\documentclass[epj]{svjour}
\usepackage{graphics}
\usepackage{float}
\usepackage{subfigure}
\usepackage{color}
\usepackage{geometry}
\usepackage{subfig}
\usepackage{caption}
\usepackage{amsfonts}
\usepackage{amsmath}
\usepackage{extarrows}
\usepackage{amssymb}
\usepackage{verbatim}
\usepackage{todonotes}
\usepackage{tikz}
\usepackage{slashed}
\usepackage{mathrsfs}
\usepackage{booktabs}
\usepackage{verbatim}
\usepackage{epsfig}
\usepackage{footnote}
 \usepackage{threeparttable}
\newcommand{\ket}[1]{|{#1}\rangle}

\newcommand{\bra}[1]{\langle{#1}|}
\newcommand{\bkt}[2]{\langle{#1}|{#2}\rangle}

\newcommand{\av}[1]{\langle#1\rangle}

\begin{document}
\title{The design  of a new fiber optic sensor for measuring linear velocity with pico meter/second sensitivity based on Weak-value amplification}
\author{Jing-Hui Huang\inst{1} \and Xue-Ying Duan\inst{2,3,4} \and Guang-Jun Wang\inst{2,3,4} \and Xiang-Yun Hu\inst{1}
\thanks{\emph{Present address:} jinghuihuang@cug.edu.cn}%
}                     
\offprints{}          
\institute{School of Institute of Geophysics and Geomatics, China University of Geosciences, Wuhan 430074, China \and School of Automation, China University of Geosciences, Wuhan 430074, China \and Hubei Key Laboratory of Advanced Control and Intelligent Automation for Complex Systems, Wuhan 430074, China \and Engineering Research Center of Intelligent Technology for Geo-Exploration, Ministry of Education}

%
\date{Received: date / Revised version: date}
%
\abstract{
We put forward a new fiber optic  sensor for measuring linear velocity with picometer/second sensitivity with Weak-value amplification based on generalized Sagnac effect [Phys. Rev. Lett.\textbf{93}, 143901(2004)].The generalized Sagnac effect was first introduced by Yao et al, which included the Sagnac effect of rotation as a special case and suggested a new fiber optic sensor for measuring linear motion with nanoscale sensitivity.
By using a different scheme to perform the Sagnac interferometer with the probe in momentum space, we have demonstrated the new weak measure protocol to detect the linear velocity by amplifying the phase shift of the generalized Sagnac effect. Given the maximum incident intensity of the initial spectrum, the detection limit of the intensity of the spectrometer, we can theoretically give the appropriate pre-selection, post-selection and other optical structures before the experiment. 
Our numerical results show our scheme with Weak-value amplification is effective and feasible to detect linear velocity with picometer/second sensitivity which is three orders of magnitude smaller than the result $\nu$=4.8 $\times$ $10^{-9}$ m/s obtained by generalized Sagnac effect with same fiber length. 
\PACS{
      {PACS-key}{discribing text of that key}   \and
      {PACS-key}{discribing text of that key}
     } 
} 
\maketitle
\section{Introduction}
\label{intro}
The monitoring of acceleration is essential for a variety of applications ranging from electronic products to scientific research. Typical accelerometer operation involves the sensitive displacement measurement of a flexibly mounted test mass, which can be realized by using capacitive, piezoelectric, tunnel-current, or optical methods\cite{2012A}. Improving the sensitivity of an accelerometer requires improving the sensitivity of detecting the linear movement of the test mass relative to the base. It is noted that the signal directly measured with high precision by these accelerometers is usually displacement rather than velocity and accelerated velocity. Hence, when measuring high-frequency signals, there will be a large error between the value of velocity obtained by numerical calculation and the real value. In that case, the picometer/second sensitivity linear velocity sensor\cite{2004Near} high precision and high sensitivity are urgently needed in many fields, such as navigation\cite{2014A} and seismology\cite{Jacopo2012Horizontal,F2013High}. 

It is believed that the Sagnac effect\cite{1913L}  has found its crucial applications in navigation as the fundamental design principle of fiber optic gyroscopes (FOGs)\cite{K2009The,2011AFORS}. The Sagnac effect\cite{1913L} shows that two counter-propagating light beams take different time intervals to travel a closed path on a rotating disk, while the light source and detector are rotating with the disk. When the disk rotates clockwise, the beam propagating clockwise takes a longer time interval than the beam propagating counterclockwise, while both beams travel the same light path in opposite directions.  The travel-time difference $\Delta t$ in a FOG can be expressed by the phase difference $\Delta \Phi = 2 \pi \Delta t c/ \lambda$, where $\lambda$ is the free space wavelength of light. On that basis, the experiments of Yao et al.\cite{2004Generalized} have discovered that any moving path contributes to the total phase difference between two counter-propagating light beams in the loop. Theirs experiments\cite{Ruyong2003Modified} using a fiber optic conveyor (FOC) showed that a segment of linearly moving glass fiber contributes to the phase difference (called "generalized Sagnac effect" in their works):
\begin{eqnarray}
\label{generalized_Sagnac_effect}
\Delta \Phi =4 \pi\oint_{l} \frac{\Vec{\nu}\ \cdot d \vec{l}}{c\lambda} 
\end{eqnarray}

Ulteriorly, they provided a new fiber linear motion sensor based on the Eq. (\ref{generalized_Sagnac_effect}), which can detect a linear velocity of $\nu =4.8 nm/s$ with $\oint_{l} d l=500 m$, $\lambda=10^{-6}m$ and the sensitivity of the FOG reaching $10^{-7}$ rad of the phase difference. However, conventional experimental schemes have difficulty detecting phase differences below $10^{-7}$ rad, in order to measure smaller velocity. 

In recent years, weak measurement has become an important area of research\cite{DLB2014Ultrasmall,2016Application,2020Approaching,2021zongsu}. In quantum mechanics, the concept of weak measurements allows for the description of a quantum system both in terms of the initial pre-selection and the final state(post-selection\cite{AAV}). The "Weak-value" was first proposed by Aharonov et al\cite{AAV}, where information is gained by weakly coupling the probe to the system. By appropriately selecting the initial and final state of the system, the measurement can be much larger than the eigenvalues of observable. Weak measurement has been utilized in metrology, and it has a variety of applications in precision detection as well as its advantage of high precision\cite{2016Application,Huang2021,2021Weak}. Dixon et al. amplified very small transverse deflections of an optical beam, then measured the angular deflection of a mirror down to 400 $\pm$ 200 frad and the linear travel of a piezo actuator down to 14 $\pm$7 fm \cite{2009Ultrasensitive}. Boyd et al. demonstrated the first realization of weak-value amplification in the azimuthal degree of freedom and had achieved effective amplification factors as large as 100\cite{PhysRevLett.112.200401}. Xu et al. implemented the phase measurement with a precision of the order of $10^{-4}$ by using a commercial light-emitting diode\cite{2013Phase}. Viza et al. achieved a velocity measurement of 400 fm/s by measuring one Doppler shifted due to a moving mirror in a Michelson interferometer\cite{2013Weak}. It's worth noting that the purpose of their research\cite{2013Weak} coincides with ours. But their measurement principle is based on the changes in the path of light in free space caused by the movement of the mirror. The optical path structure in free space is generally considered to be unstable. The requirement of the precise free-space optical alignment technique can be overcome by using the optical-fiber\cite{2003Polarization,2000Fiber}. 

In this paper, we combine the advantage of the optical fiber and the Weak-value amplification to detect linear velocity with picometer/second sensitivity. Our numerical results show our design stable optically with high sensitivity. The rest of this paper is organized in the following way. In Section 2, we present a new fiber optic sensor for measuring linear velocity with picometer/second sensitivity based on Weak-value amplification, and the numerical results are shown in Section 3. Finally, in Section 4, we give the conclusion about the work. Throughout this paper, we adopt the unit $\hbar =1$.

\begin{figure*}[t]
 \vspace{-0.2cm}
\centering
\resizebox{0.85\textwidth}{!}{%
  \includegraphics{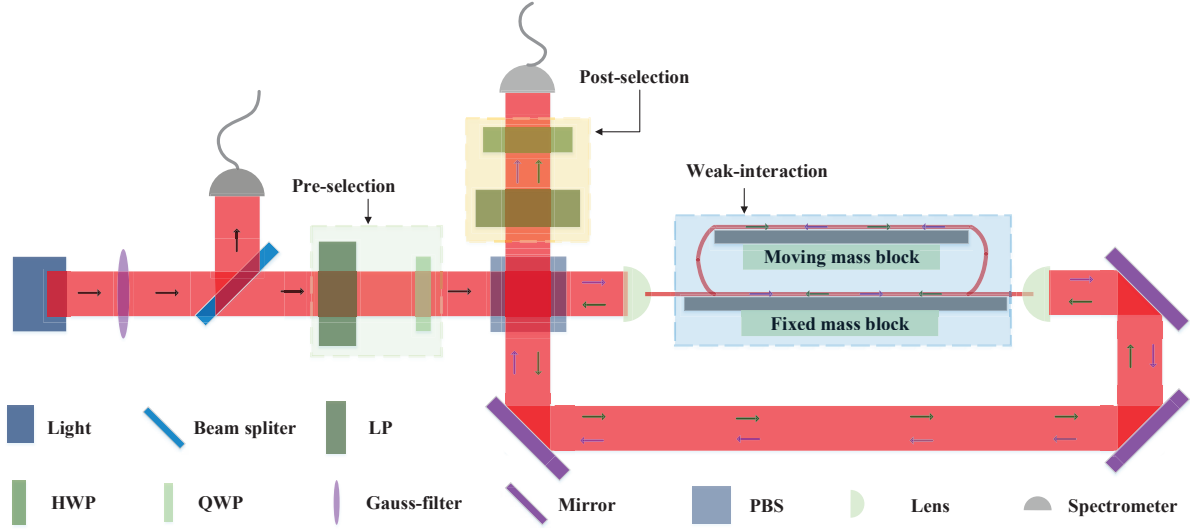}
}
\caption{Schematic of the new fiber optic sensor for measuring linear velocity with picometer/second sensitivity based on Weak-value amplification in the frequency domain. The light source is shaped by a Gauss filter(GF). BS is a beams splitter with a splitting ratio of 1$:$1000. LP is a linear polarizer for pre-selection and post-selection. HWP is the half-wave plate which induces phase shifts of $\pm \pi$ between the $\ket{H}$  and the  $\ket{V}$ components, while QWP is the quarter-wave plate that induces phase shifts of $\pm \pi / 2 $ between the$\ket{H}$  and the $\ket{V}$ components. PBS is the polarizing beam splitter which separates the light into horizontal($\ket{H}$  ) component and vertical( $\ket{V}$) component. Where $\ket{H}$ is the horizontal polarization state which circulating the loop clockwise in the Sagnac's interferometer, while  $\ket{V}$ is the vertical polarization state which circulating the same loop in a counterclockwise direction.}
\label{fig:interferometer}       
\end{figure*}

\section{Weak-value amplification for detecting weak magnetic field }
\label{sec:deltz}

In this section, we propose a new fiber optic sensor for measuring linear velocity with picometer/second sensitivity based on weak-value amplification in the frequency domain, which was reported to be superior to that in the time domain in high precision measurements\cite{PhysRevLett.105.010405}. We can detect the small velocity from the shift in the frequency domain of the probe of weak measurement, and the shift can be obtained and amplified from the weak-value. In our scheme, the optic fiber is divided into four arms, the top arm is driven by the moving mass block at a velocity $\nu$ with  picometer/second sensitivity and the bottom arm is fixed to the fixed moving mass. While moving, the two sidearms, being flexible, are kept the same shape so that the phase differences in these two sidearms cancel each other. There is no phase difference in the bottom stationary arm. Therefore, the detected phase difference is contributed solely by the motion of the top arm\cite{2004Generalized}. The other optical path structures in Fig. \ref{fig:interferometer} are designed to measure phase differences in the fiber by the amplification of weak measurement.

The weak measurement is characterized by state pre-selection, a weak perturbation, and post-selection.
We prepare the initial state $\ket{\phi_{i}}$ of the system and $\ket{\psi_{i}}$ of the probe. 
After a certain interaction between the system and the probe, we post-select a system state $\ket{\phi_{f}}$ and obtain information about a physical quantity $\hat{A}$ from the probe wave function by the weak value
\begin{eqnarray}
\label{weak_value}
{ {\rm A_{w} }:}=\frac{\bra{\phi_{f}}\hat{A}\ket{\phi_{i}}}{\bkt{\phi_{f}}{\phi_{i}}},
\end{eqnarray}
which can generally be a complex number.
More precisely, the shifts of the momentum in the probe wave function are given by the imaginary parts of the weak value Im$[{\rm A_{w} }]$. 
We can easily see from Eq. (\ref{weak_value}) that when $\ket{\phi_{i}}$ and $\ket{\phi_{f}}$ are almost orthogonal, the absolute value of the weak value can be arbitrarily large.
This leads to the weak-value amplification, as we will explain below.

The schematic diagram of the system is shown in Fig. \ref{fig:interferometer}. The incident light with a central wavelength of $\lambda_{0}$ passes through a Gaussian filter and the beam splitter. The purposes of the BS are to record the initial spectrum and to compare this spectrum with the final one. According to the previous work\cite{PhysRevLett.105.010405}, the weak measurement scheme based on circular polarizations as pre-selection and post-selection shows a higher precision than that based on linear polarizations.
Thus we select the initial polarization and the final polarization with the circular polarization. First, we prepare the initial polarization state $\ket{\phi_{0}}$ using the linear polarizer LP:
\begin{eqnarray}
\label{inter_sy_initial00}
\ket{\phi_{0}}= {\rm sin}\Big (\frac{\pi}{4}\Big)\ket{H}+ {\rm cos}\Big(\frac{\pi}{4}\Big)\ket{V}
\end{eqnarray}
where $\pi /4$ is the angle between the horizontal line and the transmission axis of LP(pre-selection); $\ket{H}$, the horizontal polarization state which circulating the loop clockwise in the Sagnac's interferometer.  $\ket{V}$, the vertical polarization state which circulating the same loop in a counterclockwise direction in the Sagnac's interferometer. Then, by introducing the QWP besides polarizers. The QWP can induce phase shifts of $ \pi /2$ between the horizontal and the vertical components. Thus the pre-selection can be presented as follow:
\begin{eqnarray}
\label{inter_sy_initial}
\ket{\phi_{i}}= {\rm sin}\Big ( \frac{\pi}{4} \Big )\ket{H}+ i {\rm cos} \Big (\frac{\pi}{4} \Big ) \ket{V}
\end{eqnarray}
The final state of the polarization is prepared by the half-wave plate HWP and the polarizer for post-selection, then the state can be expressed
\begin{eqnarray}
\label{inter_sy_final}
\ket{\phi_{f}}= i {\rm sin} \Big ( \frac{\pi}{4}+\beta \Big ) e^{i\varphi}\ket{H}
+ {\rm cos} \Big( \frac{\pi}{4}+\beta \Big ) e^{-i\varphi}\ket{V}
\end{eqnarray}
where $\beta$ is the angle between the linear polarizer of pre-selection and the linear polarizer of post-selection. The phase shift ${\varphi}$ is produced by the generalized Sagnac effect  between $\ket{H}$ and $\ket{V}$:
\begin{eqnarray}
\label{inter_sy_phase_difference}
\varphi= \frac{4\pi \nu {\rm NL} }{c \lambda_{0}}
\end{eqnarray}
where N is the number of turns of fiber optic loops, L is the length of the top arm. According to Eq. (\ref{generalized_Sagnac_effect}), it is an effective way to enhance the generalized Sagnac effect by increasing the length of the fiber loop. In Figure. \ref{fig:interferometer}, the top arm is driven by the moving mass block at a velocity $\nu$ with picometer/second sensitivity and the bottom arm is fixed to the fixed moving mass. 

In our numerical simulation, we take the initial probe wave function in Gaussian form:
\begin{eqnarray}
\label{gaussian}
|\bkt{p}{\psi_{i}}|^{2}={\Gamma}_{i}(\lambda) =I_{0}e^{-(\lambda-\lambda_{0})^{2}/W^{2}},
\end{eqnarray}
where $\lambda_{0}$  is the central wavelength of the initial spectrum and $W^{2}=(\Delta \lambda)^{2}$ is the variance of the initial spectrum. $I_{0}$ is the maximum incident intensity of the initial spectrum. Note that the wavelength $\lambda$ and the momentum $p$ of the probe has the relationship $p=2 \pi / \lambda$. In our scheme, the observable $\hat{A}$ satisfies:
\begin{eqnarray}
\label{sy_observable}
\hat{A}= \frac{1}{2}(\ket{H} \bra{H}-\ket{V} \bra{V})
\end{eqnarray}

And the weak value can be calculated by:
\begin{eqnarray}
\label{weak_value}
 {\rm A_{w} }=\frac{\bra{\phi_{f}}\hat{A}\ket{\phi_{i}}}{\bkt{\phi_{f}}{\phi_{i}}}
=\frac{{\rm sin}(\varphi){\rm sin}(\beta)+i {\rm cos}(\varphi){\rm cos}(\beta)}
{{\rm sin}(\varphi){\rm cos}(\beta)+i {\rm sin}(\beta){\rm cos}(\varphi)},
\end{eqnarray}

After weak interaction and post-selection, the absolute value squared of the probe wave function in the momentum space becomes 
\begin{eqnarray}
\label{eq_final_probe1}
|\bkt{p}{\psi_{f}}|^{2} =|\bkt{\phi_{f}}{\phi_{i}}|^{2}
 \times e^{2pg {\rm Im}({\rm A_{w} })} |\bkt{p}{\psi_{i}}|^{2}
\end{eqnarray}
where $|\bkt{\phi_{f}}{\phi_{i}}|^{2}$ is the probability to pass the post-selection.
\begin{eqnarray}
\label{eq_final_probe2}
|\bkt{\phi_{f}}{\phi_{i}}|^{2} ={\rm sin}^{2}(\varphi){\rm cos}^{2}(\beta) 
+ {\rm sin}^{2}(\beta){\rm cos}^{2}(\varphi)
\end{eqnarray}

An important practical advantage of weak measurements lies in measuring the imaginary part of the weak value, which shifts a variable conjugate to the one which is normally affected by the relevant interaction\cite{PhysRevA.76.044103}. Furthermore, the shifted variable is corresponding the center wavelength $\lambda_{0}$. The observer $\hat{A}$ will perform a readout of the probe, by measuring its shift, thus obtaining velocity information about the system\cite{PhysRevLett.105.010405}. In our weak measurement protocol, the observer $\hat{A}$ and the shift of the center wavelength $\lambda_{0}$ has the relationship $\Delta\av{\hat{p}}= 2g W^{2} {\rm Im} {\rm A_{w} }$ with $g=2 \pi / p_{0}$ \cite{1990Properties,2018Optical,Li:16}, $p_{0}$ corresponds to the center wavelength $\lambda_{0}=2 \pi / p_{0}$. Finally, because the weak value $  {\rm A_{w} }$ is the function of measured the velocity $v$, the velocity $v$ and the imagine part of the weak value $ {\rm Im} {\rm A_{w} }$
can be obtained from the shift of the center wavelength $\lambda_{0}$:
\begin{eqnarray}
\label{inter_delt_lambda}
\delta \lambda_{0}=-\frac{4 \pi (\Delta \lambda)^{2}}{\lambda_{0}}  {\rm Im} {\rm A_{w} }
\end{eqnarray}
where the imaginary part $ {\rm Im} {\rm A_{w} }$ of the weak value can be calculated from Eq. (\ref{weak_value})
\begin{eqnarray}
\label{weak_value_img}
 {\rm Im} {\rm A_{w} }
=\frac{{\rm sin}(\varphi){\rm cos}(\varphi)({\rm cos}^{2}(\beta)-{\rm sin}^{2}(\beta))}
{{\rm sin}^{2}(\varphi){\rm cos}^{2}(\beta)+{\rm sin}^{2}(\beta){\rm cos}^{2}(\varphi)},
\end{eqnarray}

Finally, the relationship of the shift of the central wavelength $\delta \lambda_{0}$ and the linear velocity can be obtained from Eq. (\Ref{inter_delt_lambda}). It is noting that the imagined part of the weak value can be efficiently amplified by choosing the appropriate angle $\beta$ between the pre-selection state and the post-selection state.
In this paper, the results of our numerical simulation will be shown in the next section.

\section{Numerical result and discussion}
In this part, we simulate the experimental setup for measuring linear velocity with a picometer/second sensitivity based on the weak value amplification. That is, by choosing the appropriate post-selected state and the initial probe in momentum space, we can calculate the weak value and the shift of the center wavelength of the probe before the experiment. In order to perform the numerical simulation, the parameters of the main optical components in our scheme in Fig. \ref{fig:interferometer} shall meet the following requirements: 

\textbf{Light}: High-power broadband light sources is suitable for our scheme, and the light normally adopt Superluminescent diodes($\lambda_{0}$ = 840 nm) in these works\cite{Li:16,2018Optical} for weak measurement in momentum space. The higher the intensity of the initial light, the greater the successful probability of success of post-selection\cite{PhysRevLett.105.010405}. Note that the shift of the central wavelength $\delta \lambda_{0}$ is proportional to the spectral width $\Delta \lambda$. Thus, the wider the bandwidth of the light source we choose, the greater the amplification effect of the weak value. Therefore, we chose the initial probe wave function (\ref{gaussian}) with $\lambda_{0}=840$ nm and $W^{2}=(\Delta \lambda)^{2}=150$ nm$^2$.

\begin{figure}[htp!]
	\centering
	\centerline{\includegraphics[scale=0.54,angle=0]{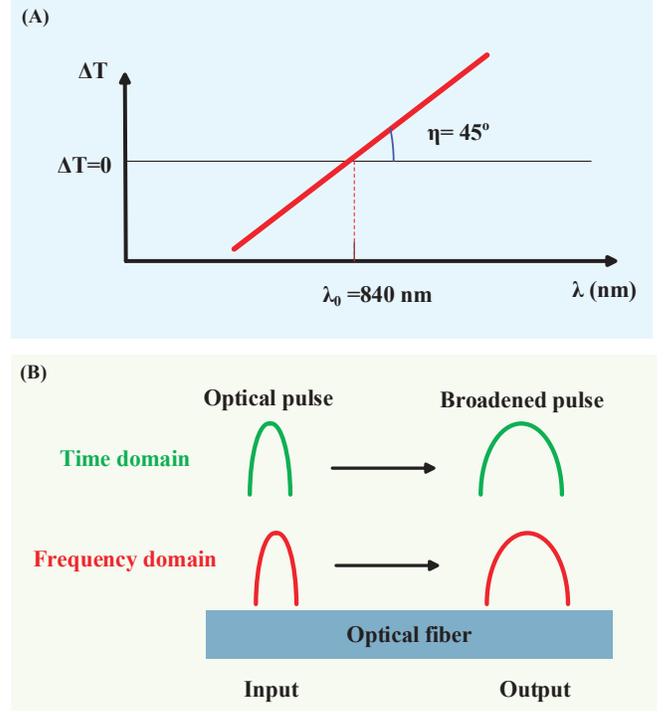}}
\vspace*{0mm} \caption{\label{Fig:fiber-character}Optical fiber for the weak measurement. (A) the dispersion cure $\Delta T- \lambda$ and (B) transmission characteristics of the fiber we chose.}
\end{figure}

\textbf{Fiber}: In principle, the optical fiber length should be as long as possible. The generalized Sagnac effect (\ref{inter_sy_phase_difference}) shows that increasing the length of the fiber can increase the phase difference caused by the generalized Sagnac effect and thus improve the sensitivity of the measurement. On the other hand, normally, dispersion is the primary cause of limitation on the optical signal transmission bandwidth through an optical fiber\cite{1071236}. Dispersion is the spreading out of a light pulse in time as it propagates down the fiber. Dispersion in optical fiber includes model dispersion, material dispersion and waveguide dispersion. In our work, material dispersion may greatly affect weak measurements. Material dispersion is caused by the wavelength dependence of the refractive index on the fiber core material, while waveguide dispersion occurs due to the dependence of the mode propagation constant on the fiber parameters and signal wavelength. However, In our scheme, the material dispersion can be effectively eliminated by selecting special optical fibers: (i) selecting the fiber with the lowest dispersion coefficient; (ii)as is shown in Fig. \ref{Fig:fiber-character}(A), finding the fiber with a controlled dispersion slope, where the dispersion curve is symmetric about the central wavelength. And the related research can be found in the literature\cite{1071236,Photonic_devices,10.1109/68.789717}. In that case, although the spectrum will widen, the central wavelength $\delta \lambda_{0}$ of the spectrum will not be shifted by the dispersion effect(the progress is shown in Fig. \ref{Fig:fiber-character}(B)). In particular, in order to compare our work to the traditional scheme\cite{2004Generalized}, we take the total length of the top arm of optical fiber NL = 500 m, where the length is the same length in the work\cite{2004Generalized}. 

\begin{figure}[htp!]
	\centering
	\centerline{\includegraphics[scale=0.76,angle=0]{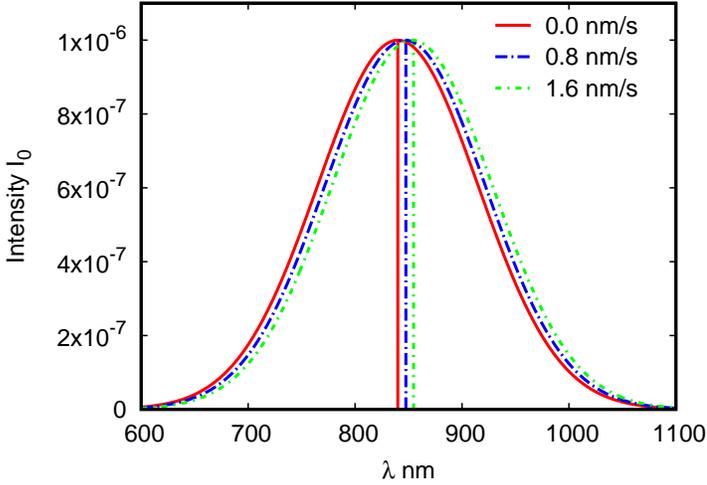}}
\vspace*{0mm} \caption{\label{result_numberical1}The central wavelength shifts of our simulation experiment with post-selection angle $\beta$ = 0.001 rad. The vertical lines represent the central wavelengths of the corresponding Gaussian spectrum. }
\end{figure}

\begin{figure}[htp!]
	\centering
	\centerline{\includegraphics[scale=0.72,angle=0]{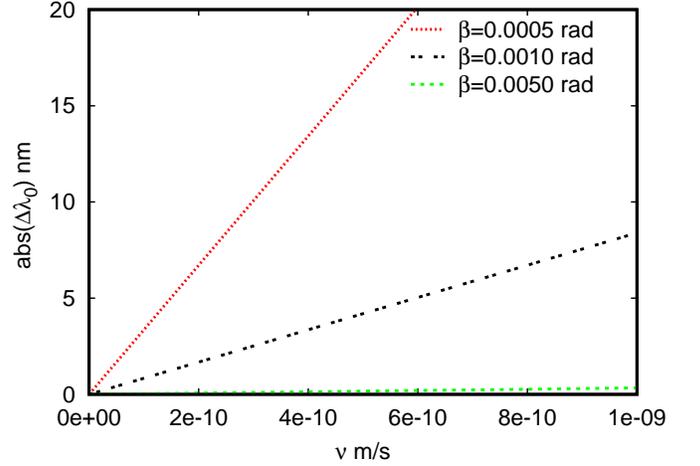}}
\vspace*{0mm} \caption{\label{result_numberical2}The shifts of the central wavelength dependence of linear velocity with different post-selection angles.}
\end{figure}

Then, by using the Eqs. (\ref{eq_final_probe1}) and (\ref{inter_delt_lambda}), we obtain the final probe function and the shifts of the center wavelength with different values of the post-selection angle $\beta$. The simulation results are shown in Fig. \ref{result_numberical1}, Fig. \ref{result_numberical2} and Table. \ref{tab:1}.

\begin{table}[t]
\centering
\caption{Parameters and the numerical values of our obtained results: NL presents the total length of the top arm;
$\beta $  is corresponding to the post-selection ; k=$d| \Delta \lambda_{0})|/d \nu$ is the sensitivity of our scheme and the absolute value of the slope in the Fig. \ref{result_numberical2}.}
\label{tab:1}       
\begin{tabular}{lllll}
\toprule
 NL(m) & $\beta $ (rad)&  k (nm/m/s) &  $|\bkt{\phi_{f}}{\phi_{i}}|^{2}$ \\
\noalign{\smallskip}\hline\noalign{\smallskip}
500&0.0050 & 3.4 $\times10^{8}$ & 2.5 $\times10^{-5}$\\
500&0.0010 & 5.4 $\times10^{9}$ & 1.0 $\times10^{-6}$\\
500&0.0005 & 3.4 $\times10^{10}$ & 2.5 $\times10^{-7}$\\
\toprule\\
\end{tabular}
\vspace*{-0.6cm}  
\end{table}

Fig. \ref{result_numberical1} shows the shifts of the Gaussian spectrum with post-selection angle $\beta$ = 0.001 rad at different linear velocities. By fitting each central wavelength of the Gaussian spectrum at different linear velocity, we obtain the shifts of the Gaussian spectrum due to the change of the linear velocity.
More specifically, the shift of the central wavelength increases as the linear velocity increases.

Fig. \ref{result_numberical2} and the Table. \ref{tab:1} show the sensitivity of our scheme with different post-selection angles $\beta$, which corresponds to the slope of the curve. Our numerical results show the smaller the $\beta$ is, the larger the amplification(corresponding to the slope k) is. On the other hand, the smaller $\beta$ leads to the lower probability $|\bkt{\phi_{f}}{\phi_{i}}|^{2}$ to detect the post-selection. Therefore, the value of $\beta$ cannot be infinitesimally small due to the measurement limit of the detection instrument and the low signal-to-noise ratio. It is worth noting that it is possible to detect the signal at the value of reachable signal-to-noise ratio of -68 dB\cite{1999The}, due to the great advantage of using chaotic oscillators in weak signal detection. In that case, the main condition that limits the accuracy of our system measurements is the resolution of the spectrometer $\Delta \lambda_{min}$. 

Compared with the traditional fiber optic conveyor based on generalized Sagnac effect\cite{Ruyong2003Modified,2004Generalized} to measure linear velocity, our scheme base on Weak-value amplification is fundamentally different in the principle of measurement. In traditional fiber optic interference, the phase shift $\varphi$ (\ref{inter_sy_phase_difference}) induced by generalized Sagnac effect is normally detected by the change of light intensity $I=A[1+ {\rm cos} (\varphi) ]$, the sensitivity of a fiber-optic gyroscope can be $10^{-7}$ rad of the phase difference by detecting the change of light intensity. In the traditional fiber optic gyroscope, the stability of optical power is affected by the fluctuation of optical power, temperature and so on\cite{1993Optical}.  However, the detected signal in our design of the new fiber optic sensor based on weak-value amplification is not affected by changes in light intensity. Besides, the detected velocity can be infinitely magnified due to the advantage of weak measurement. The equation (\ref{weak_value}) and  equation (\ref{inter_delt_lambda}) indicates that the weak value $\rm A_w$, as well as the detected velocity $v$ can be amplified by choosing ${\bkt{\phi_{f}}{\phi_{i}}} \rightarrow 0$. Note that the main factor limiting the sensitivity of our measurements is the spectral resolution of the measuring spectrometer. To sum up, the above discussion can explain why the sensitivity is improved in our new fiber optic sensor based on weak measurement.

Usually, the resolution of the spectrometer $\Delta \lambda_{min}$ can reach 0.02 nm, our scheme can detect a linear velocity of $\nu=\Delta \lambda_{min} / k$=3.7 $\times$ $10^{-12}$ m/s with NL=500m, $\beta$=0.001 rad. The result which is picoscale velocity and is three orders of magnitude smaller than the result $\nu$=4.8 $\times$ $10^{-9}$ m/s obtained by "generalized Sagnac effect"\cite{2004Generalized} by considering same fiber length. This is due to the advantage of weak-value amplification of weak measurements.

We have made the first step towards the numerical study of Weak-value amplification for detecting linear velocity with picometer/second sensitivity. Our numerical results show our scheme can effectively measure linear velocity with picometer/second sensitivity.

\section{Conclusion}
In conclusion, we use weak-value amplification to measuring linear velocity with picometer/second sensitivity based on the generalized Sagnac effect. By choosing the appropriate pre-selection, post-selection, and the initial probe in the momentum space, we obtain the relationship between the shifts of the center wavelength and linear velocity. Our numerical results show that the weak-value amplification can outperform conventional measurement in the presence of detector saturation. Comparing with detecting the change of light intensity\cite{2004Generalized}, our measurement scheme can have the remarkable advantage of avoiding the affected by changes in optical power. Besides, our numerical results show our scheme with Weak-value amplification is effective and feasible to detect linear velocity with picometer/second sensitivity which is three orders of magnitude smaller than the result $\nu$=4.8 $\times$ $10^{-9}$ m/s obtained by generalized Sagnac effect. 

Before performing specific experiments, our numerical results have confirmed that our scheme can effectively measure linear velocity with picometer/second sensitivity. Besides, in order to detect the weaker linear velocity, it is effective and feasible to increase the maximum incident intensity $I_{0}$ of the initial spectrum, improve the measurement precision of the spectrometer, and increase the length of the fiber loop. On the other hand, at the given the maximum incident intensity $I_{0}$ of the initial spectrum, the detection limit of the intensity of the spectrometer and the accuracy of detecting linear velocity, we can theoretically give the appropriate post-selection, pre-selection and others optical structure before the experiment. In addition, the relevant optical experiments are taking in progress.
%
%
\section{Authors contributions}
Jing-Hui Huang and Xue-Ying Duan collected the literature and wrote the article. Guang-Jun Wang and Xiang-Yun Hu revised the article. Jing-Hui Huang designed the study. Xue-Ying Duan prepared figures and tables. All the authors were involved in the preparation of the manuscript. All the authors have read and approved the final manuscript.

\section*{Acknowledgements}
This study is financially supported by the National Key Research and Development Program of China (Grant No. 2018YFC1503705) and the Fundamental Research Funds for National Universities, China University of Geosciences(Wuhan) (Grant No. G1323519204).
%
%

\bibliographystyle{ws-mpla}
\bibliography{reference}

\end{document}